\documentclass[sigconf]{acmart}
\AtBeginDocument{%
  \providecommand\BibTeX{{%
    \normalfont B\kern-0.5em{\scshape i\kern-0.25em b}\kern-0.8em\TeX}}}

\usepackage{graphicx}
\usepackage{subcaption}

\renewcommand\footnotetextcopyrightpermission[1]{} 
\usepackage{cleveref}

\usepackage{caption}
\usepackage{enumitem}

\renewcommand\footnotetextcopyrightpermission[1]{} 
\begin{document}

\title{RobQuNNs: A Methodology for Robust Quanvolutional Neural Networks against Adversarial Attacks}


\author{Walid El Maouaki\textsuperscript{1}, 
Alberto Marchisio\textsuperscript{2,3}, 
Taoufik Said\textsuperscript{1},
Muhammad Shafique\textsuperscript{2,3},
and Mohamed Bennai\textsuperscript{1}
}

\affiliation{%
  \institution{\textsuperscript{1}Quantum Physics and Magnetism Team, LPMC, Faculty of Sciences Ben M’Sik, Hassan II University of Casablanca, Morocco}
  \country{\textsuperscript{2}eBrain Lab, Division of Engineering, New York University Abu Dhabi (NYUAD), Abu Dhabi, UAE\\
  \textsuperscript{3}Center for Quantum and Topological Systems (CQTS), NYUAD Research Institute, NYUAD, Abu Dhabi, UAE}
  }
\email{walid.elmaouaki-etu@etu.univh2c.ma, alberto.marchisio@nyu.edu, taoufik.said@univh2c.ma,}
\email{muhammad.shafique@nyu.edu, mohamed.bennai@univh2c.ma}

\renewcommand{\shortauthors}{El Maouaki, et al.}

\begin{abstract}
Recent advancements in quantum computing have led to the emergence of hybrid quantum neural networks, such as Quanvolutional Neural Networks (QuNNs), which integrate quantum and classical layers. While the susceptibility of classical neural networks to adversarial attacks is well-documented, the impact on QuNNs remains less understood. This study introduces RobQuNN, a new methodology to enhance the robustness of QuNNs against adversarial attacks, utilizing quantum circuit expressibility and entanglement capability alongside different adversarial strategies. Additionally, the study investigates the transferability of adversarial examples between classical and quantum models using RobQuNN, enhancing our understanding of cross-model vulnerabilities and pointing to new directions in quantum cybersecurity. The findings reveal that QuNNs exhibit up to 60\% higher robustness compared to classical networks for the MNIST dataset, particularly at low levels of perturbation. This underscores the potential of quantum approaches in improving security defenses. In addition, RobQuNN revealed that QuNN does not exhibit enhanced resistance or susceptibility to cross-model adversarial examples regardless of the quantum circuit architecture.

\end{abstract}






\maketitle
\pagestyle{plain}

\section{Introduction}

As the world increasingly relies on artificial intelligence (AI) for safety-critical applications in healthcare~\cite{yu2018artificial}, autonomous driving~\cite{muhammad2020deep}, and security systems~\cite{zhang2022artificial}, the robustness of AI systems against adversarial attacks becomes paramount~\cite{hamon2020robustness}. Before integrating Quantum Machine Learning (QML) into such high-risk areas, it is vital to develop models that can withstand adversarial environments and malicious manipulations. This need is especially urgent for Quanvolutional Neural Networks (QuNNs), whose robustness remains underexplored compared to their classical counterparts. Studies referenced in~\cite{akter2023exploring} and~\cite{wendlinger2024comparative} highlight notable adversarial weaknesses in both classical and quantum neural networks. The findings suggest that, despite quantum neural networks demonstrating enhanced resilience and inherent robustness stemming from their unique structures, they remain susceptible to significant adversarial attacks. However, these claims lack thorough empirical validation regarding the properties of the quantum circuits. Additionally, a recent work in~\cite{liu2020vulnerability} has explored the relationship between Hilbert space dimensionality and adversarial vulnerability in quantum neural networks, demonstrating that higher dimensions can lead to decreased robustness. However, a key question remains: can properties of the Hilbert space itself be leveraged to improve the robustness of QuNNs?  Moreover, the work in~\cite{lu2020quantum} demonstrated that gradient-based adversarial attacks are transferable from classical neural networks to quantum neural networks. However, their study did not explore the reverse transferability, where attacks originate from quantum models and impact classical systems. Additionally, the influence of quantum model architecture on this transferability was not examined in their work.

Our work addresses existing gaps by introducing the RobQuNN methodology, which explores whether specific features of the Hilbert space can enhance robustness. The RobQuNN methodology discovers that the expressibility and entanglement capability of the Hilbert space, controllable through the configuration of quantum circuit gates, leads to greater robustness in QuNNs. This methodology focuses on the role of the quantum circuit architecture (Ansatze) in improving resilience against adversarial attacks. We systematically investigate how different Ansatze architectures, reflecting varying levels of Hilbert space expressibility and entanglement capability, influence the model's behavior under various adversarial conditions, including cross-model adversarial attacks, which is a critical area that has been insufficiently explored in the existing literature. Our exploration not only reveals compelling findings regarding the impact of Hilbert space entanglement capability on QuNNs' robustness and cross-model adversarial resilience but also highlights the significant lack of systematic studies on these models under diverse adversarial scenarios.
By delving into how quantum-specific features can influence security and robustness in hybrid quantum-classical systems, our study contributes to a more comprehensive understanding of QuNNs' vulnerabilities and paves the way for developing more secure quantum-based AI applications.

In this paper, we present \textit{RobQuNN}, a methodology for robust QuNNs against adversarial attacks. Our contributions to this paper are:\vspace{-0.5em}

\begin{itemize}
    \item We design the RobQuNN methodology to robustify QuNNs against different types of white box adversarial attacks through the quantum circuit architecture. (\textbf{\Cref{sec:methodology}})
    \item We compare the robustness of QuNNs to classical CNNs against various types of adversarial attacks. (\textbf{\Cref{subsec:comparison_Classical_Quantum}})
    \item We evaluated the vulnerability of QuNN model against black box attacks uisng RobQuNN. (\textbf{\Cref{Transferibility}})
\end{itemize}


\section{Background}
\subsection{Convolutional and Quanvolutional neural networks}
Convolutional Neural Networks (CNNs)~\cite{li2021survey}, ~\cite{alzubaidi2021review}, have become an essential component of current image identification tasks due to their unique architecture, which is capable of capturing spatial structure in data. A CNN architecture typically consists of a sequence of layers: convolutional layers, pooling layers, and fully connected layers. The convolutional layer, the fundamental building block, applies a set of learnable filters, or kernels, to the input image to generate feature maps. Mathematically, the convolution operation is defined as $(X * K)[i, j]=\sum_m \sum_n X[i+m, j+n] K[m, n]$, where $X$ is the input image, $K$ is the kernel, and $*$ denotes the convolution operation. This process extracts features such as edges, textures, and shapes. Pooling layers, such as max-pooling or average-pooling, are then used to downsample the feature maps, reducing the spatial dimensions and the computational load while retaining the most significant features. \textit{In this study, we concentrate specifically on the convolutional layer in our models and deliberately omit the pooling layer. This decision allows us to delve deeper into the convolutional layer's unique contributions to feature extraction and model robustness}. 

QuNNs represent an innovative hybrid architecture that leverages quantum computing to enhance classical CNNs by incorporating quantum circuits to process data~\cite{henderson2020quanvolutional}. In a QuNN, the convolutional layers are replaced by quantum circuits, specifically designed to exploit quantum superposition and entanglement, enabling the extraction of complex features that might be challenging for classical CNNs. The quantum convolutional layer, also known as the quanvolutional layer, functions similarly to a classical convolutional layer but utilizes quantum circuits to process image patches. For an $m\times m$ input image, an $n\times n$ kernel, and a stride of $2$, the quanvolutional layer processes data using a $n^2$ qubits quantum circuit. The dimensionality of the feature maps is directly influenced by the number of qubits used, with each qubit contributing to a distinct channel in the output. QuNN process data as follows:

\textbf{Extracting patches:} Each $n \times n$ patch $P_{i, j}$ of the image $X$ is defined by:
$$
P_{i, j}=\{X[i+k, j+l] \mid 0 \leq k<n, 0 \leq l<n\}
$$
where $i$ and $j$ denote the coordinates of the top-left corner of the patch.

\textbf{Encoding patches into quantum states:} Each $n \times n$ patch $P_{i, j}$ is encoded into a quantum state $\left|\psi_{i, j}\right\rangle$. One common encoding method is angle encoding, where it maps each pixel value from the patch to the corresponding qubit using a rotation gate:
$$
\left|\psi_{i, j}\right\rangle=U_{\mathrm{enc}}\left(P_{i, j}\right)|0\rangle^{\otimes n^2}
$$
where $U_{\text {enc }}$ is the unitary operator for encoding the patch into a quantum state over $n^2$ qubits.

\textbf{Applying the quanvolutional filter (quantum kernel):} The encoded quantum state $\left|\psi_{i, j}\right\rangle$ is processed by an $n^2$-qubit operation $U(\theta)$ which acts as an $n \times n$ kernel:
$$
\left|\phi_{i, j}\right\rangle=U(\theta)\left|\psi_{i, j}\right\rangle
$$
Here, $U(\theta)$ is a unitary transformation designed to perform the quantum equivalent of convolution. $U(\theta)$ is a parameterized quantum circuit (PQC) which can vary depending on the gate selection and entanglement pattern chosen.

\textbf{Measurement:} Each qubit in the quantum state $\left|\phi_{i, j}\right\rangle$ is measured, and its expectation value contributes to a specific channel in the feature map. For an $n \times n$ kernel with $n^2$ qubits, the expectation value of each qubit $k$ is computed as:
$$
M_{i, j, k}=\left\langle\phi_{i, j}\left|Z_k\right| \phi_{i, j}\right\rangle
$$
where $Z_k$ is the Pauli-Z operator for the $k$-th qubit. This measurement yields a scalar value that corresponds to one pixel in the channel $k$.

\textbf{Constructing the feature map:} The feature map is constructed by applying these steps to each $n \times n$ patch, with a stride of 2. The resulting feature map $F$ consists of multiple channels $F_k$, each corresponding to one of the qubits. Formally, the feature map can be described as:
$$
F_k[i, j]=M_{i, j, k}
$$
where $k$ ranges from 1 to $n^2$, representing the different channels, and $i$ and $j$ vary in steps of 2 , i.e., $i, j \in\{0,2,4, \ldots, m-n\}$.

\textbf{Quanvolutional operation:} The quanvolutional layer's operation can be succinctly represented as:
$$F_k[i, j] = \left\langle U(\theta) U_{\mathrm{enc}}\left(P_{i, j}\right) \left| 0 \right\rangle^{\otimes n^2} \middle| Z_k \middle| U(\theta) U_{\mathrm{enc}}\left(P_{i, j}\right) \left| 0 \right\rangle^{\otimes n^2} \right\rangle
$$

\begin{figure*}[h]
    \centering
    \includegraphics[width=\linewidth]{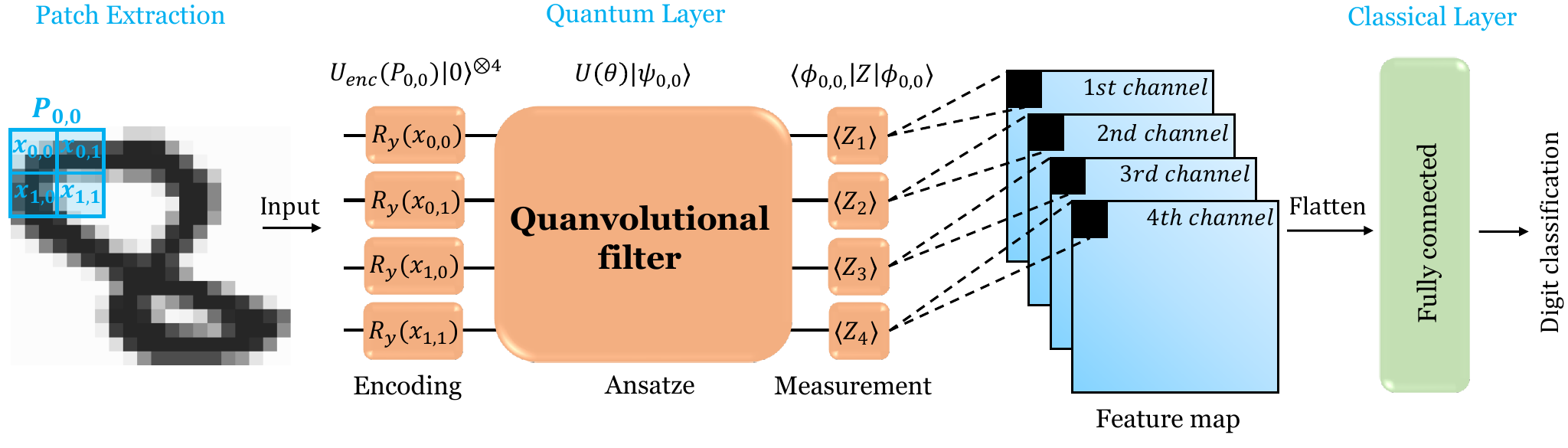}
    \caption{
    Illustration of a $2\times 2$ kernel QuNN applied to the digit $'8'$ from the MNIST dataset. The procedure commences with the extraction of a $2 \times 2$ patch from the input image. Each pixel value within this patch is then encoded into a quantum state using a rotation gate around the $y$-axis. Subsequently, a Quanvolutional filter applies a parameterized quantum circuit (Ansatze) to the quantum states. The output is measured to produce feature maps across multiple channels. These feature maps are then flattened and processed through a fully connected classical layer to perform digit classification.}
    \label{fig:QNN_architecture}
\end{figure*}

Finally, the feature maps produced by either the CNN or the QuNN are flattened and passed through one or more fully connected layers, which perform classification using the learned features. Refer to figure \ref{fig:QNN_architecture} for an illustration of a QuNN applied to the digit $'8'$ from the MNIST dataset.

\subsection{Quantum machine learning vulnerability}
The susceptibility of quantum classifiers to adversarial attacks has gained notable attention. Research such as~\cite{liu2020vulnerability} and~\cite{lu2020quantum} showcased that QML models, much like classical models, can be vulnerable to adversarial examples. They have shown that even when these quantum classifiers are implemented on noisy quantum devices, they can still be deceived by these subtle alterations to the input. As a defense strategy, one proposed technique is adversarial training, which involves including these adversarial examples during the training process to improve the model's resilience.

Building on this empirical studies, there has been progress in theoretical approaches to enhancing the adversarial robustness of QML models. For instance, this study~\cite{georgiou2024adversarial} established new information-theoretic upper bounds on the generalization error of adversarially trained quantum classifiers, establishing a crucial link between adversarial perturbations and increased sample complexity. Another significant contribution in this field employed quantum many-body physics principles to offer provable defenses for QML models~\cite{dowling2024adversarial}. This research demonstrated that quantum classifiers can inherently resist low perturbations, local attacks under non-scrambling conditions, and universal adversarial attacks in scenarios characterized by quantum chaos, fundamentally strengthening the security of QML algorithms against sophisticated adversarial attacks. These insights help illuminate the underlying dynamics of QML models under adversarial conditions, offering a pathway toward the development of more robust QML systems. 

Moreover, incorporating quantum phenomena into defense strategies is emerging as a novel method to protect QML models. Recent studies propose using quantum noise and other unique quantum properties to enhance adversarial robustness, potentially improving theoretical limits on adversarial errors~\cite{du2021quantum},~\cite{huang2023certified},~\cite{gong2024enhancing}.

\section{RobQuNNs Methodology}
\label{sec:methodology}

Our methodology explores new avenues for robustifying QuNNs. RobQuNNs use the Hilbert space or quantum circuit expressibility and entanglement capability to enhance the QuNN robustness against various adversarial attacks. In the realm of quantum machine learning, the expressiveness of quantum neural networks plays a vital role in performing various tasks. Typically, the greater the expressiveness and entanglement ability of quantum neural networks, the more effectively they can represent the solution space and establish more distinct decision boundaries~\cite{sim2019expressibility}. The expressibility of the QuNN quanvolutional layer depends on the number of layer repetitions, the number and selection of single-qubit gates, and the axes around which the qubits are rotated. This includes, for example, rotations around the $x, y$, and $z$ axes (denoted by $R_x, R_y$, and $R_z$, respectively). Whereas the entanglement capability is determined by its two-qubit gates selection, such as $CNOT, CR_x, CR_y, CR_z, etc$, as well as the entanglement pattern between qubits~\cite{sim2019expressibility}. In this work, for a fair comparison across all Ansatz architectures, we use the same rotation gates and the same number of layers in the circuit, by using a single-layer repetition and incorporating the same rotation gates around the three axes in our circuit. We concentrate on varying the entanglement capability of the quanvolutional layer in the QuNN by varying the quantum circuit's entanglement from low to high. Our goal is to assess how different levels of entanglement influence data representation within the quantum circuit, affect the construction of the decision boundary, and impact the QuNN's resilience against various adversarial attacks and across a range of perturbation strengths. In our work, we explored various entanglement patterns, including architectures that follow popular nearest-neighbor and all-to-all entanglement topologies, as well as random and no entanglement architectures. These architectures exhibit varying degrees of entanglement capability. Additionally, we examined how adversarial examples transfer between QuNNs and CNNs, assessing quantum circuit entanglement capability effectiveness in cross-model attacks and highlighting security challenges to guide the development of robust defense strategies in mixed-model environments.

\subsection{Ansatzes architectures}
We developed five Ansatze architectures with different entanglement capabilities, and they are as follows:

\begin{itemize}
    \item No entangled qubit: This Ansatz consists only of single-qubit gates applied to each qubit. It includes rotation gates around the $x, y, \text{and}, z$ axes. There is 0 entanglement operation. This Ansatz is expressed as: 
    \begin{equation}
        U_{noent}=\left(\bigotimes_{0}^n \text{Rot}\left(\alpha, \beta, \gamma\right)\right)
    \end{equation}
    where $\text{Rot}\left(\alpha, \beta, \gamma\right)=R_z(\alpha)R_y(\beta)R_z(\gamma)$
    
    \item  Full qubit Entanglement: This Ansatz follows the all-to-all topology with $ZZ$ two-qubit gates, connecting every qubit in the circuit. It requires $n(n-1)/2$ entanglement operations, where $n$ is the number of qubits. Additionally, it includes single qubit rotation gates. The Ansatze expression is:
    \begin{equation}
    U_{\text{full}} = O^{\text{FullEnt}}_{ZZ}\left(\bigotimes_{0}^n \text{Rot}\left(\alpha, \beta, \gamma\right)\right)
    \end{equation}
    
    \begin{equation}
    O^{\text{FullEnt}}_{ZZ} := \prod_{p=1}^{n-1} \prod_{q=p+1}^n \mathbf{G}_{p, q} \;\;\;\text{where}\;\;\; \mathbf{G}_{p, q} = e^{-i \varphi Z_p Z_q}
    \end{equation}

    \item Linear qubit entanglement: This Ansatze follows the nearest-neighbor topology and it connects each neighbor pair of qubits using the ZZ two-qubit gates. This entanglement pattern requires $n+1$ operations where n is the number of qubits. This Ansatze is expressed as:
    \begin{equation}
    U_{\text{linear}} = O^{\text{LinearEnt}}_{ZZ}\left(\bigotimes_{0}^n \text{Rot}\left(\alpha, \beta, \gamma\right)\right)
    \end{equation}
    
    \begin{equation}
    O^{\text{LinearEnt}}_{ZZ} := \prod_{p=1}^{n-1} \mathbf{G}_{p, p+1} \;\;\;\text{where}\;\;\; \mathbf{G}_{p, p+1} = e^{-i \varphi Z_p Z_{p+1}}
    \end{equation}

    \item Star qubit entanglement: This Ansatz follows a star-shaped entanglement topology, with $ZZ$ two-qubit gates, connecting one central qubit to all other qubits. For $n$ qubits, it requires $n-1$ entanglement operations. In our case, the entanglement is centered on the first qubit. Additionally, it includes single qubit rotation gates. The Ansatze expression is: 
    \begin{equation}
    U_{\text{star}} = O^{\text{StarEnt}}_{ZZ}\left(\bigotimes_{0}^n \text{Rot}\left(\alpha, \beta, \gamma\right)\right)
    \end{equation}
    
    \begin{equation}
    O^{\text{StarEnt}}_{ZZ} := \prod_{p=1}^{n-1} \mathbf{G}_{1, p+1} \;\;\;\text{where}\;\;\; \mathbf{G}_{1, p+1} = e^{-i \varphi Z_1 Z_{p+1}}
    \end{equation}

    \item Random architecture: This Ansatze features random gate selection and includes only a single entanglement operation.
    
\end{itemize}


\subsection{Adversarial attacks on QuNNs and CNNs} \label{ADVattack}
Given a machine learning model, whether classical or quantum, we train it iteratively until it converges to a local optimal solution, achieving a satisfactory separation of the dataset.

\textit{White Box Adversarial Attacks}: In a white-box adversarial attack scenario, the adversary has full knowledge of the model, including its architecture, parameters, and gradients. This knowledge is exploited to craft adversarial examples that aim to mislead CNN into making incorrect classifications. The following methods are employed to generate these adversarial examples:

\begin{enumerate}
    \item Fast Gradient Sign Method (FGSM) \cite{goodfellow2014explaining}:
 is a gradient based attack that perturbs the input image $x$ by adding a small perturbation $\epsilon$ in the direction of the gradient of the loss function $J(\theta, x, y)$ with respect to the input, where $\theta$ represents the model parameters, and $y$ is the true label. The adversarial example $x^{\prime}$ is computed as:
$$
x^{\prime}=x+\epsilon \cdot \operatorname{sign}\left(\nabla_x J(\theta, x, y)\right)
$$

Here, $\epsilon$ is a small scalar value that controls the strengths of the perturbation.
    \item Projected Gradient Descent (PGD) \cite{madry2018towards}:
 is an iterative extension of FGSM. It applies multiple iterations of small FGSM like steps and projecting the perturbed image back into the feasible input space. The adversarial example after $k$ iterations is given by:
$$
x_{k+1}=\operatorname{Proj}_\epsilon\left(x_k+\alpha \cdot \operatorname{sign}\left(\nabla_x J\left(\theta, x_k, y\right)\right)\right)
$$
where $\alpha$ is the step size, and $\operatorname{Proj}_\epsilon$ denotes the projection operator that ensures $x_{k+1}$ remains within the $\epsilon$-ball around the original input $x$.

    \item Momentum Iterative Method (MIM) \cite{dong2018boosting}:
 is an enhancement of PGD that incorporates a momentum term to stabilize the direction of the perturbation, thereby improving the attack's effectiveness. The momentum-based gradient is calculated as:
$$
g_{k+1}=\mu \cdot g_k+\frac{\nabla_x J\left(\theta, x_k, y\right)}{\left\|\nabla_x J\left(\theta, x_k, y\right)\right\|_1}
$$
where $g$ is the accumulated gradient and $\mu$ is the momentum factor. The adversarial example is updated as: 
$$
x_{k+1}=\operatorname{Proj}_\epsilon\left(x_k+\alpha \cdot \operatorname{sign}\left(g_{k+1}\right)\right)
$$
\end{enumerate}

\textit{Black box attacks}: Cross-model transferability in black-box adversarial attacks is the phenomenon where white-box adversarial examples, created to trick one model, are transferred to fool other models, even if those models have different architectures.

\section{Results and Discussion}

\begin{figure}
    \centering
    \includegraphics[width=0.9\linewidth]{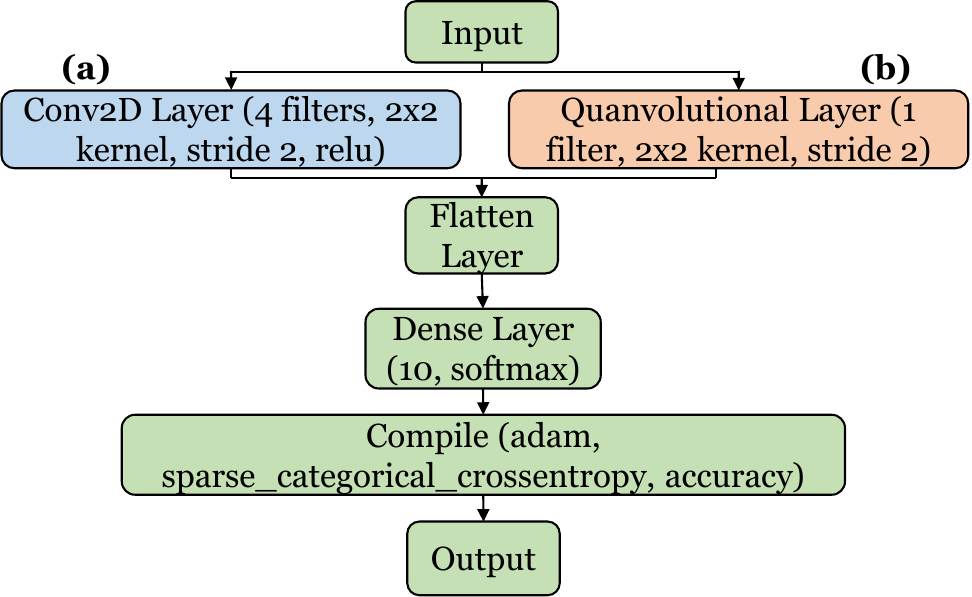}
    \caption{Diagram showing (a) a CNN with a convolutional layer, and (b) a QuNN with a quantum convolutional layer, both for the MNIST dataset. Post-flattening, the data is sent to the softmax layer, followed by model optimization and performance assessment.}
    \label{fig:Flows}
\end{figure}

\subsection{Experimental Setting}
The adversarial robustness of QuNN and CNN was assessed using the MNIST dataset~\cite{mnist}, consisting of 28x28 pixel grayscale images, against three adversarial attack methods: FGSM, PGD, and MIM, each at varying levels of perturbation strength. See Figure \ref{fig:Flows} for an overview of the quantum and classical model architectures. 

The CNN model utilizes a classical Conv2D layer with 4 filters and uses the ReLU activation function to introduce non-linearity. In contrast, QuNNs employ a quantum convolutional (Quanvolutional) layer, which uses a single filter corresponding to one quantum circuit layer and a 2x2 kernel representing 4 qubits in the circuit, with a stride of 2. Both models produce feature maps with dimensions of 14x14 pixels and 4 channels.
After convolution, each model flattens the resulting feature maps into a one-dimensional vector, which is then fed into a dense layer with 10 neurons. This layer uses the softmax activation function to output a probability distribution across 10 classes (digits 0-9). Each model is compiled using the Adam optimizer, with sparse categorical cross-entropy as the loss function and accuracy as the performance metric.
For this study, in both models, the convolutional layers act as fixed feature extractors for both classical and quantum models, meaning their parameters are non-trainable. However, the fully connected layer is trainable, enabling model optimization.

The models are trained for 30 epochs with a batch size of four, and a learning rate of $0.001$. We use these trained models for white-box attacks to generate adversarial examples using different attack methods: FGSM, PGD, and MIM. The examples are generated using the TensorFlow Cleverhans library, which computes the gradient of the loss function on the image pixels and updates the adversarial examples as described in subsection \ref{ADVattack}. We vary the perturbation strength from low to high relative to each attack type, and create a set of adversarial examples for each level of perturbation. Lastly, the performance of the trained models is assessed using the adversarial images across the range of epsilon values, which produces plots of accuracy against epsilon.

\subsection{White Box Adversarial Attacks on CNNs and QuNNs}
\label{subsec:comparison_Classical_Quantum}

\begin{figure}
    \centering
    \includegraphics[width=\linewidth]{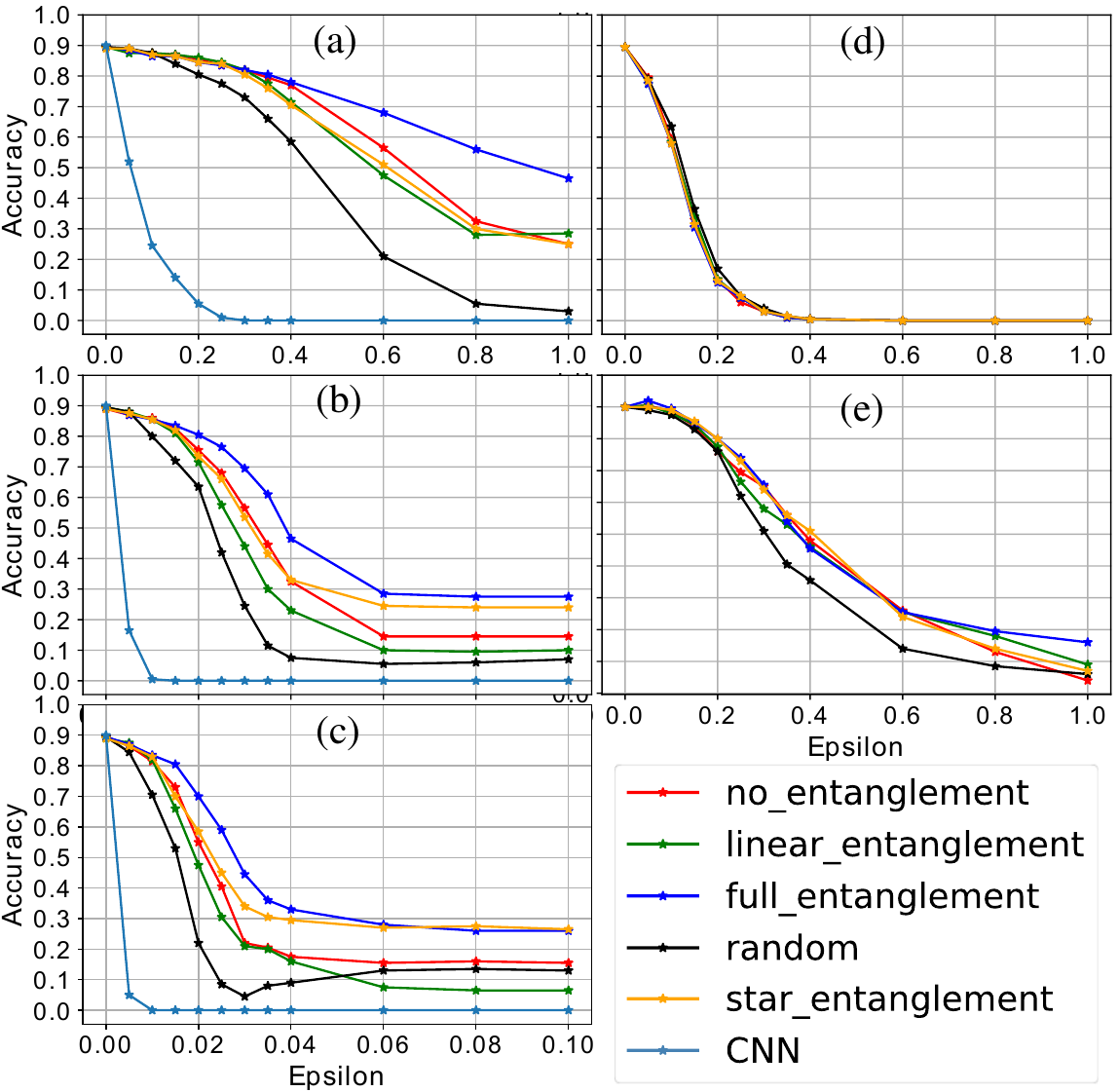}
    \caption{Robustness Evaluation of CNN and QuNN: Left — Robustness of classical and quantum models against a) FGSM, b) PGD, and c) MIM adversarial attacks at varying perturbation strengths for the MNIST dataset. Right — Cross-model adversarial attack robustness: (d) Adversarial examples from classical models transferred to the quantum model, (e) Adversarial examples from the quantum model transferred to classical models.}
    \label{QuNNvsCNN}
\end{figure}

Compared to a classical CNN, the QuNN model was consistently more resistant to adversarial attacks. While both models' accuracy decreased with stronger attacks (FGSM Figure \ref{QuNNvsCNN}a, PGD Figure \ref{QuNNvsCNN}b, MIM \ref{QuNNvsCNN}c), the QuNN's decline was slower. Notably, the QuNN maintained higher accuracy across a range of attack strengths, demonstrating better robustness under harsher conditions. 
Within the QuNN models themselves, different architectures showed varying resilience. The ZZ full architecture performed best on the MNIST dataset, showing strong resilience and effective preservation of the data robust features, while Random architecture was the least robust. Additionally, at low epsilon values, we observe that architectures with similar entanglement patterns tend to exhibit comparable behavior. 
This is demonstrated in both the ZZ full and star entanglement configurations, where a common feature of these architectures is that in both patterns, the first qubit is entangled with all other qubits.

  


\subsection{Transferability of Adversarial Examples} \label{Transferibility}
In the conducted experiments Figure \ref{QuNNvsCNN}(d,e), adversarial examples crafted using the FGSM demonstrated a notably high level of transferability between classical and quantum models. This was evident across multiple tests and circuit architecture variations within QuNNs.

\textbf{Classical to Quantum, Figure \ref{QuNNvsCNN}d:} When white box adversarial examples generated by the CNN were applied to the quantum models, all quantum Ansatzes exhibited similar declines in performance. This uniform response suggests that adversarial examples developed in the classical setting retain their effectiveness when transferred to quantum models, exploiting similar weaknesses despite the underlying differences in model complexity.

\textbf{Quantum to Classical, Figure \ref{QuNNvsCNN}e:} Conversely, when white box adversarial examples generated by the quantum models were employed against the classical CNN, a consistent pattern of effectiveness was also observed. The classical model's performance was significantly compromised by adversarial inputs devised across different quantum Ansatze. However, the classical model preserved a slightly increased robustness compared to the classical transferability to the quantum model.

These findings collectively highlight a critical aspect of adversarial example utility and risk, showing that adversarial perturbations are capable of crossing the computational boundaries between quantum models and classical models with minimal loss in attack efficacy.


\subsection{Discussion}
The adversarial robustness comparison between QuNN and CNN using the MNIST dataset under FGSM, PGD, and MIM attacks showcases the superior resilience of QuNNs, particularly with the ZZ full architecture. This architecture consistently outperformed others in resisting various intensities of adversarial perturbations, affirming the potential of specific quantum Ansatze entanglement capability to enhance security in environments prone to white-box adversarial threats. The results indicate that increasing the connections between qubits in the quanvolutional circuit enhances its ability to extract robust features from images. The varied performance across different QuNN architectures underscores the critical role of architectural choices in achieving enhanced adversarial robustness in quantum-inspired models.

On the other hand, the results in Section \ref{Transferibility} reveal that adversarial examples demonstrate high transferability between classical and quantum models, underscoring substantial security implications for machine learning systems in an era increasingly influenced by quantum computing. Despite the quantum model's robust architecture, it is susceptible to adversarial examples crafted using classical models, which are more effective and capable of deceiving the quantum model. This indicates the classical model's low resistance to the generation of adversarial examples. Conversely, the classical model exhibits moderate robustness against adversarial examples generated by the quantum model, reflecting the resistance of the QuNN against generating effective adversarial examples due to its inherent quantum properties. This suggests that the QuNN model, even under a white box attack, effectively retains meaningful and robust features \cite{west2023towards}. This demonstrates its efficiency in feature extraction, which could explain the moderate performance of the classical model. However, in contrast, classical models struggle to preserve robust features from the data, leading to a reduction in QuNN’s robustness.

Additionally, adversarial examples crafted by classical models retain their effectiveness when transferred to the quantum model, regardless of the entanglement strategy used. This consistency suggests that, in the case of black-box attacks, the vulnerabilities exploited are fundamental to the data or features, rather than being specific to the model architecture. However, the opposite is observed in white-box attack scenarios. Overall, quantum models do not exhibit enhanced resistance or susceptibility to cross-model adversarial examples, indicating that the core vulnerabilities are common across different quantum circuit architectures.

\section{Conclusion}
We developed RobQuNN methodology that explores the quantum circuit design to increase QuNN resilience. We examined how the design of QuNN Ansatze affects their security against adversarial attacks. In a white-box attack environment, we tested five different entanglement configurations within the circuits and found that ZZ full entanglement, and Ansatze with closer entanglement pattern, offered the most robust defense on the MNIST dataset. This suggests that, unlike classical models, QuNNs' security depends heavily on their specific design, making this study the first to explore this critical link.

Furthermore, by applying the RobQuNN framework, we broadened the range of adversarial threats to encompass cross-model adversarial attacks—examining the transferability of attacks between classical and quantum models to assess the resilience of the QuNN model. We discovered that in these scenarios, unlike in white-box attacks, the QuNN Ansatze architecture does not enhance the model's robustness. This indicates that for such attacks, the design of the QuNN Ansatze is less critical, and we should explore other model features to improve its adversarial resistance. Although QuNNs are robust against direct attacks, they remain susceptible to adversarial examples devised from different models, underscoring the importance of comprehensive adversarial defense strategies.

\begin{acks}
This work was supported in parts by the NYUAD Center for Quantum and Topological Systems (CQTS), funded by Tamkeen under the NYUAD Research Institute grant CG008, and the NYUAD Center for Cyber Security (CCS), funded by Tamkeen under the NYUAD Research Institute Award G1104.
\end{acks}

\bibliographystyle{ieeetr}
\bibliography{main.bib}

\end{document}